\documentclass{cmip-l}%
\usepackage{amsmath}
\usepackage{amsfonts}
\usepackage{amssymb}
\usepackage[dvips]{graphicx}%
\providecommand{\U}[1]{\protect\rule{.1in}{.1in}}
\newtheorem{theorem}{Theorem}
\newtheorem{acknowledgement}[theorem]{Acknowledgement}

\newtheorem{conjecture}[theorem]{Conjecture}

\begin{document}
\begin{titlepage}
\vspace{.3cm} \vspace{1cm}
\begin{center}
\baselineskip=16pt \centerline{\large\bf  Noncommutative Geometry as the Key to Unlock the Secrets of Space-Time } \vspace{2truecm} \centerline{\large\bf Ali H.Chamseddine\  } \vspace{.5truecm}
\emph{\centerline{Physics Department, American University of Beirut, Lebanon}}
\emph{\centerline{email: chams@aub.edu.lb}}
\end{center}
\vspace{2cm}
\begin{center}
{\bf Abstract}
\end{center}
I give a summary of the progress made on using the elegant construction of
Alain Connes noncommutaive geometry to explore the nature of space-time at very
high energies. In particular I show that by making very few natural  and  weak assumptions about the structure of
the noncommutative space, one can deduce
the structure of all fundamental interactions at low energies.
\end{titlepage}

\section{\bigskip Introduction}

This article is dedicated to Alain Connes on the occasion of his 60th
birthday. I have come to know Alain well during my first visit to IHES in
1996. I\ was immediately overwhelmed with his brilliance and the overflow of
his ideas, and within a short time started to collaborate with him on the
interface of noncommutative geometry, his invention, and the ideas of
unification in theoretical physics. This collaboration has been very fruitful,
and we have come to appreciate the mysterious links between geometry and
physics. Many problems remain, but I am optimistic that the challenge of
finding a quantum theory of gravity using the geometric tools that Alain
developed, is within reach. At the personal level, I\ discovered that Alain is
a very warm person, full of life, and has fantastic sense of humor. I\ am
proud of his friendship.

What I\ will present here, is a summary of a forthcoming long article in
collaboration with Alain, which hopefully will appear in the near future,
where a self contained exposition of the methods of noncommutative geometry
applied to particle physics are explained in a language accessible to
physicists \cite{Prepare}. A good part of this forthcoming article will
elaborate and build on the results that were first obtained with the crucial
input of the collaboration with Marcolli \cite{mc2}. In addition, the
introduction present in a recent paper \cite{Uncanny} can be used to help
introduce the reader to the general philosophy of our program. Our aim is to
provide enough material to help students and young researchers who wish to
learn about this promising direction of research.

The laws of physics at low energies are well encoded by the action functional
which is the sum of the Einstein-Hilbert action and that of the standard
model. These two parts have different properties, the first being dependent on
the geometry of the underlying manifold $\left(  M,g\right)  $ where $g$ is
the metric, while the other is governed by internal symmetries of a gauge
group $G$ which can be well described using the language of vector bundles.
The underlying symmetries are also different. General relativity is governed
by diffeomorphism invariance $\left(  \text{outer automorphisms of }\left(
M,g\right)  \right)  $ while gauge symmetries are based on local gauge
invariance (inner automorphisms). Thus the natural group of invariance is the
semi direct product
\[
G=\mathit{U}\rtimes\hbox{Diff}\left(  M\right)
\]
where
\[
\mathit{U}=C^{\infty}\left(  M,U(1)\times SU(2)\times SU(3)\right)  .
\]
It is possible to trace back the failure of finding a unified theory of all
interactions including quantum gravity to the difference between these two
kinds of symmetries. In addition, there are many unanswered questions within
the established formulation of the standard model. For example, the following
questions have no compelling answer: Why the gauge group is specifically given
by { $U(1)\times SU(2)\times SU(3)$ ? Why the fermions occupy the particular
representations that they do? Why there are three families and why there are
}$16$ fundamental fermions per family? What is the theoretical origin of the
Higgs mechanism and spontaneous breakdown of gauge symmetries? What is the
Higgs mass and how to explain all the fermionic masses? These are only few of
the questions that have to be answered by the ultimate unified theory of all
interactions. We shall attempt to answer some of these questions taking as a
starting point the following observations. At energies well below the Planck
scale
\[
{M_{P}=\sqrt{\frac{1}{8\pi G}}\equiv\frac{1}{\kappa}=2.43\times10^{18}%
\hbox{ Gev}}%
\]
gravity can be safely considered as a classical theory. But as energies
approach the Planck scale one expects the quantum nature of space-time to
reveal itself, and for the Einstein-Hilbert action to become an approximation
of some deformed theory. In addition the other three forces must be unified
with gravity in such a way that all interactions will correspond to one
underlying symmetry. One thus would expect that the nature of space-time, and
thus of geometry, would change at Planckian energies, in such a way that at
lower energies, one recovers the above picture of diffeomorphism and internal
gauge symmetries. It is not realistic to guess the exact properties of
space-time at Planckian energies and to make directly an extrapolation of $17$
orders of magnitude from our present energies. We are therefore led to take an
indirect approach where we search for the hidden structure in the functional
of gravity coupled to the standard model at present energies. To do this we
shall make a basic conjecture which we will take as a starting point:

\begin{conjecture}
{\emph{At some energy level, space-time is the product of a continuous
four-dimensional manifold times a discrete space F}. }
\end{conjecture}

{The aim then is to find supporting evidence for this conjecture. Once this is
done the next step would be to find the true geometry at Planckian energies,
for which this product in turn is a limit. }

This is the minimal extension where no new extra dimensions are assumed. The
task now is to determine with {minimal input} the properties of the discrete
space F, and construct the associated physical theory. Remarkably, we\ will
show that this information will allow us to determine the hidden structure of
space-time, and answer some, but not all (so far) of the questions posed above.

\section{{\emph{A Brief Summary of Alain Connes NCG}}}

The basic idea is based on physics. The modern way of measuring distances is
spectral. The unit of distance is taken as the wavelength of atomic spectra.
To adapt this geometrically{ the notion of real variable which one takes as a
function $f$ on a set $X$ where {$f:X\rightarrow R$ has to be replaced. } This
is now taken to be a self adjoint operator in a Hilbert space as in quantum
mechanics. The space $X$ \ is described by the algebra $\mathcal{A}$ of
coordinates which is represented as operators in a fixed Hilbert space
$\mathcal{H}$. }The geometry of the noncommutative space is determined in
terms of the spectral data $\left(  {\mathcal{A},\mathcal{H},\mathcal{D}%
,J\ ,\gamma}\right)  $ .\ A { real, even spectral triple } is defined by
\cite{Book}, \cite{CoSM}

\begin{itemize}
\item {$\mathcal{A}$ is an associative algebra with unit 1 and involution
{$\ast$}.}

\item {$\mathcal{H}$ is a complex Hilbert space carrying a faithful
representation {$\pi$} of the algebra.}

\item $D${ is a self-adjoint operator on {$\mathcal{H}$} with the resolvent
$\left(  D-\lambda\right)  ^{-1},\lambda\in\mathbf{R}$ of $D$ compact.}

\item {$J$ is an anti--unitary operator on {$\mathcal{H}$}, which is a real
structure (charge conjugation.)}

\item {$\gamma$ is a unitary operator on {$\mathcal{H}$}, the chirality.}
\end{itemize}

We require the following axioms to hold:

\begin{itemize}
\item {$J^{2}=$}$\varepsilon$ , ($\varepsilon=1$ in zero dimensions and
$\varepsilon=-1$ in 4 dimensions).

\item {$[a,b^{o}]=0$} for all {$a,b\in\mathcal{A}$, where $b^{o}=Jb^{\ast
}J^{-1}.$ This is the zeroth order condition. This is needed to define the
right action on elements of $\mathcal{H}$} : $\zeta b=b^{o}\zeta,$ and is a
statement that left action and right action commute.

\item {$DJ=\varepsilon^{\prime}JD,\quad J\gamma=\varepsilon^{\prime\prime
}\gamma J,\quad D\gamma=-\gamma D$} where {$\varepsilon,\varepsilon^{\prime
},\varepsilon^{\prime\prime}\in\left\{  -1,1\right\}  .$} The reality
conditions resemble the conditions governing the existence of Majorana (real) fermions.

\item {$[[D,a],b^{o}]=0$} for all {$a,b\in\mathcal{A}$}. This is the first
order condition.

\item {$\gamma^{2}=1$} and {$[\gamma,a]=0$} for all {$a\in\mathcal{A}$}. These
properties allow the decomposition {$\mathcal{H}=\mathcal{H}_{L}%
\oplus\mathcal{H}_{R}$}.

\item {$\mathcal{H}$ is endowed with {$\mathcal{A}$} bimodule structure
$a\zeta b=ab^{o}\zeta.$ }

\item The notion of dimension is governed by growth of eigenvalues, and may be
{fractal or complex}.
\end{itemize}

$\mathcal{A}$ has a well defined unitary group
\[
\mathcal{U}=\left\{  u\in\mathcal{A};\quad u\,u^{\ast}=u^{\ast}u=1\right\}  .
\]
{The natural adjoint action of $\mathcal{U}$ on $\mathcal{H}$ is given by
$\zeta\rightarrow u\zeta u^{\ast}=u\,J\,u\,J^{\ast}\zeta\quad\forall\zeta
\in\mathcal{H}.$} Then{%
\[
\left\langle \zeta,D\zeta\right\rangle
\]
} is not invariant under the above transformation: {%
\[
\left(  u\,J\,u\,J^{\ast}\right)  D\left(  u\,J\,u\,J^{\ast}\right)  ^{\ast
}=D+u\left[  D,u^{\ast}\right]  +J\left(  u\left[  D,u^{\ast}\right]  \right)
J^{\ast}.
\]
However, }the action $\left\langle \zeta,D_{A}\zeta\right\rangle $ is
invariant, where
\[
{D_{A}=D+A+\varepsilon^{\prime}JAJ^{-1},\quad A=%
{\displaystyle\sum\limits_{i}}
a^{i}\left[  D,b^{i}\right]  }%
\]
and $A=A^{\ast}$ is self-adjoint. \ This is similar to the appearance of the
interaction term for the photon with the electrons
\[
{i\overline{\psi}\gamma^{\mu}\partial_{\mu}\psi\rightarrow i\overline{\psi
}\gamma^{\mu}\left(  \partial_{\mu}+ieA_{\mu}\right)  \psi}%
\]
to maintain invariance under the variations {$\psi\rightarrow e^{ie\alpha
\left(  x\right)  }\psi.$}

The properties listed above of the anti-linear isometry {$J:\mathcal{H}%
\rightarrow\mathcal{H}$ are characteristic of} a real structure of
{$KO$-dimension $n\in\mathbb{Z}/8$} on the spectral triple {$(\mathcal{A}%
,\mathcal{H},D).$} The numbers {$\varepsilon,\varepsilon^{\prime}%
,\varepsilon^{\prime\prime}\in\{-1,1\}$} are a function of $n$ mod $8$ given by

\begin{center}
{%
\begin{tabular}
[c]{|c|rrrrrrrr|}\hline
\textbf{n } & 0 & 1 & 2 & 3 & 4 & 5 & 6 & 7\\\hline\hline
$\varepsilon$ & 1 & 1 & -1 & -1 & -1 & -1 & 1 & 1\\
$\varepsilon^{\prime}$ & 1 & -1 & 1 & 1 & 1 & -1 & 1 & 1\\
$\varepsilon^{\prime\prime}$ & 1 &  & -1 &  & 1 &  & -1 & \\\hline
\end{tabular}
}
\end{center}

\bigskip We take the algebra {$\mathcal{A}$} to be given by a tensor product
which geometrically corresponds to a product space. The spectral geometry of
{$\mathcal{A}$} is given by the product rule {$\mathcal{A}=C^{\infty}\left(
M\right)  \otimes\mathcal{A}_{F}$} \ where the algebra {$\mathcal{A}_{F}$} is
finite dimensional, and {
\[
\mathcal{H}=L^{2}\left(  M,S\right)  \otimes\mathcal{H}_{F},\quad
D=D_{M}\otimes1+\gamma_{5}\otimes D_{F},
\]
} where {$L^{2}\left(  M,S\right)  $} is the Hilbert space of {$L^{2}$}
spinors, and {$D_{M}$} is the Dirac operator of the Levi-Civita spin
connection on $M$, {${D_{M}=\gamma^{\mu}\left(  \partial_{\mu}+\omega_{\mu
}\right)  .}$} The Hilbert space $\mathcal{H}_{F}$ is taken to include the
physical fermions. The chirality operator is {$\gamma=\gamma_{5}\otimes
\gamma_{F}$ and the reality operator is }$J=C\otimes J_{F}$, where $C$ is the
charge conjugation matrix.

In order to avoid the fermion doubling problem where the fermions $\zeta
,\zeta^{c},\zeta^{\ast},\zeta^{c\ast}$, $\zeta\in\mathcal{H},$ should not be
all independent, it was shown that the finite dimensional space must be taken
to be of K-theoretic dimension $6$ \cite{AC}, \cite{Barrett}, where in this
case {$\left(  \varepsilon,\varepsilon^{\prime},\varepsilon"\right)
=(1,1,-1)$ }$\left(  \hbox{so as to impose the condition }J\zeta=\zeta\right)
.$ This makes the total K-theoretic dimension of the noncommutative space to
be $10$ and would allow to impose the reality (Majorana) condition and the
Weyl condition simultaneously in the Minkowskian continued form, a situation
very familiar in ten-dimensional supersymmetry. In the Euclidean version, the
use of the $J$ \ in the fermionic action, would give for the chiral fermions
in the path integral, a {Pfaffian} instead of determinant \cite{AC}, and will
thus cut the fermionic degrees of freedom by a factor of 2. In other words, in
order to have the fermionic sector free of the fermionic doubling problem we
\ must make the choice {
\[
J_{F}^{\,2}=1,\qquad J_{F}D_{F}=D_{F}J_{F},\qquad J_{F}\,\gamma_{F}%
=-\gamma_{F}J_{F}.
\]
} In what follows we will restrict our attention to determination of the
finite algebra, and will omit the subscript $F$.

\section{{Classification of Finite Noncommutative Spaces}}

There are two main constraints on the algebra from the axioms of
noncommutative geometry. We first look for involutive algebras ${\mathcal{A}}$
of operators in {${\mathcal{H}}$} such that,{
\[
\lbrack a,b^{0}]=0\,,\quad\forall\,a,b\in{\mathcal{A}}\,,
\]
} where for any operator {$a$ in ${\mathcal{H}}$, $a^{0}=Ja^{\ast}J^{\,-1}$.}
This is called the order zero condition. We shall assume that the
representations of ${\mathcal{A}}$ and $J$ \ in ${\mathcal{H}}$ are
\emph{irreducible.}

The classification of the irreducible triplets {$\left(  \mathcal{A}%
,\mathcal{H},J\right)  $ leads to the } following theorem \cite{beggar},
\cite{SM}:

\begin{theorem}
{\emph{The center }$Z\left(  \mathcal{A}_{\mathbb{C}}\right)  $\emph{ is }%
}$\mathbb{C}${\emph{ or }}$\mathbb{C}${$\oplus$}$\mathbb{C}${\emph{.}}
\end{theorem}

{If {the center $Z\left(  \mathcal{A}_{\mathbb{C}}\right)  $ is $\mathbb{C}$}
then we can state the following theorem: }

\begin{theorem}
{ \-\textit{Let \ }${\mathcal{H}}$\textit{ be a Hilbert space of dimension
}$n$\textit{. Then an irreducible solution with }$Z\left(  \mathcal{A}%
_{\mathbb{C}}\right)  =$\textit{ }}$\mathbb{C}${\textit{ exists iff }$n=k^{2}%
$\textit{ is a square. It is given by }$A_{\mathbb{C}}=M_{k}\left(
\mathbb{C}\right)  $\textit{ acting by left multiplication on itself and
anti-linear involution }%
\[
J\left(  x\right)  =x^{\ast},\quad\forall x\in M_{k}\left(  \mathbb{C}\right)
.
\]
}
\end{theorem}

\bigskip For {{$\mathcal{A}_{\mathbb{C}}=M_{k}\left(  \mathbb{C}\right)  $ we
have $\mathcal{A=}M_{k}\left(  \mathbb{C}\right)  ,$ $M_{k}\left(
\mathbb{R}\right)  $} or {$M_{a}\left(  \mathbb{H}\right)  $} for even
{$k=2a,$ where $\mathbb{H}$} is the field of quaternions \cite{bour}.} These
correspond respectively to the unitary, orthogonal and symplectic case.

{If {the center $Z\left(  \mathcal{A}_{\mathbb{C}}\right)  $ is
$\mathbb{C\oplus C}$ then} we can state the theorem:}

\begin{theorem}
{ \-\textit{Let \ }${\mathcal{H}}$\textit{ be a Hilbert space of dimension
}$n$\textit{. Then an irreducible solution with }$Z\left(  \mathcal{A}%
_{\mathbb{C}}\right)  =$\textit{ }}$\mathbb{C}${$\oplus$}$\mathbb{C}${\textit{
exists iff }$n=2k^{2}$\textit{ is twice a square. It is given by
}$A_{\mathbb{C}}=M_{k}\left(  \mathbb{C}\right)  \oplus M_{k}\left(
\mathbb{C}\right)  $\textit{ acting by left multiplication on itself and
anti-linear involution }%
\[
J\left(  x,y\right)  =\left(  y^{\ast},x^{\ast}\right)  ,\quad\forall x,y\in
M_{k}\left(  \mathbb{C}\right)  .
\]
}
\end{theorem}

With each of the {$M_{k}\left(  \mathbb{C}\right)  $} in {$\mathcal{A}%
_{\mathbb{C}}$} we can have the three possibilities {$M_{k}\left(
\mathbb{C}\right)  ,$}{ $M_{k}\left(  \mathbb{R}\right)  ,$} or {$M_{a}\left(
\mathbb{H}\right)  ,$} where {$k=2a$}. At this point we make the
\textit{hypothesis} that we are in the {\textquotedblleft
symplectic--unitary"} case, thus restricting the algebra {$\mathcal{A}$} to
the form
\[
{\mathcal{A}=M_{a}\left(  \mathbb{H}\right)  \oplus M_{k}\left(
\mathbb{C}\right)  ,\ \qquad k=2a.}%
\]
The dimension of the Hilbert space is {$n=2k^{2},$ however, because of the
reality condition,} these correspond to {$k^{2}$} fundamental fermions , where
{$k=2a$} is an even integer. The first possible value for $k$ is $2$
corresponding to a Hilbert space of four fermions and an algebra
{$\mathcal{A}=\mathbb{H}\oplus M_{2}\left(  \mathbb{C}\right)  $.} This is
ruled out because it does not allow to impose grading on the algebra. It is
also ruled out by the existence of quarks. The next possible value for {$k$ is
$4$} thus predicting the number of fermions to be {$16.$}

In the {$Z\left(  \mathcal{A}_{\mathbb{C}}\right)  =$ $\mathbb{C}$} case, one
can show that it is not possible to have the finite space to be of
\ K-theoretic dimension $6$ consistent with the relation $J\gamma=-\gamma J$
\cite{beggar}. We therefore can proceed directly to the second case.

One then takes the grading $\gamma$ of $\mathcal{H}$ so that the K-theoretic
dimension of the finite space is $6$ and this is consistent with the condition
{$J\,\gamma=-\gamma J.$} It is given by{
\[
\gamma\left(  \zeta,\eta\right)  =\left(  \gamma\zeta,-\gamma\eta\right)  .
\]
} This grading breaks the algebra {$\mathcal{A=}$ $M_{2}\left(  \mathbb{H}%
\right)  \oplus M_{4}\left(  \mathbb{C}\right)  $}, which is non trivially
graded only for the {$M_{2}\left(  \mathbb{H}\right)  $ }component, to its
even part:{
\[
\mathcal{A}^{\hbox{ev}}=\hbox{ }\mathbb{H}_{R}\oplus\mathbb{H}_{L}\oplus
M_{4}\left(  \mathbb{C}\right)  \,.
\]
}

The Dirac operator must connect the two pieces non-trivially, and therefore
must satisfy {
\[
\left[  D,Z\left(  \mathcal{A}\right)  \right]  \neq\left\{  0\right\}  .
\]
} The physical meaning of this constraint, is to allow some of the fermions to
acquire Majorana masses, realizing the see-saw mechanism, and thus connecting
the fermions to their conjugates.

We have to look for subalgebras {$\mathcal{A}_{F}\subset\mathcal{A}%
^{\hbox{ev}},$} the even part of the algebra $\mathcal{A}$, for which
{$[[D,a],b^{0}]=0,\quad\forall\,a,b\in\mathcal{A}_{F}$.} We can state the theorem:

\begin{theorem}
{\emph{Up to automorphisms of }$A^{\hbox{ev}},$\emph{ there exists a unique
involutive subalgebra }$A_{F}\subset A^{\hbox{ev}}$\emph{ of maximal dimension
admitting off-diagonal Dirac operators}{
\begin{align*}
\mathcal{A}_{F}  &  =\left\{  \left(  \lambda\oplus\overline{\lambda}\right)
\oplus q,\,\lambda\oplus m\,|\lambda\in\mathbb{C},\quad q\in\mathbb{H},\quad
m\in M_{3}\left(  \mathbb{C}\right)  \right\} \\
&  \subset\mathbb{H}\oplus\mathbb{H}\oplus M_{4}\left(  \mathbb{C}\right)  .
\end{align*}
} It is \emph{isomorphic to }}$\mathbb{C}${$\oplus\mathbb{H}\oplus
M_{3}\left(  \mathbb{C}\right)  $\emph{. }}
\end{theorem}

\section{{Tensor Notation}}

It is helpful to write the results obtained about the standard model using
tensor notation. The Dirac action must take the form {%
\[
\Psi_{M}^{\ast}D_{M}^{N}\Psi_{N}%
\]
} where {$\Psi_{M}=\left(
\begin{array}
[c]{c}%
\psi_{A}\\
\psi_{A^{\prime}}%
\end{array}
\right)  $} and we have denoted {$\psi_{A^{\prime}}=\psi_{A}^{c}$}, the
conjugate spinor. We start with the algebra{
\[
\mathcal{A}=M_{4}\left(  \mathbb{C}\right)  \oplus M_{4}\left(  \mathbb{C}%
\right)
\]
and }denote the spinors by $\psi_{A}=\psi_{\alpha I}$, $A=\alpha I,\quad
\alpha=1,\cdots,4,\quad I=1,\cdots,4,$\ and thus $D_{A}^{B}=D_{\alpha
I}^{\beta J}.$ The Dirac operator takes the form{
\[
D=\left(
\begin{array}
[c]{cc}%
D_{A}^{B} & D_{A}^{B^{^{\prime}}}\\
D_{A^{^{\prime}}}^{B} & D_{A^{^{\prime}}}^{B^{^{\prime}}}%
\end{array}
\right)  ,
\]
} where $A^{\prime}=\alpha^{\prime}I^{\prime},\quad\alpha^{\prime}=1^{\prime
},\cdots,4^{\prime},\quad I^{\prime}=1^{\prime},\cdots,4^{\prime}$, and
$D_{A^{^{\prime}}}^{B^{^{\prime}}}=\overline{D}_{A}^{B}$, $D_{A^{^{\prime}}%
}^{B}=\overline{D}_{A}^{B^{^{\prime}}}$ and overbar denotes complex conjugation.

Elements of the algebra $\mathcal{A}$ are matrices $a_{M}^{N}$ of the special
form: {
\[
a=\left(
\begin{array}
[c]{cc}%
X_{\alpha}^{\beta}\delta_{I}^{J} & 0\\
0 & \delta_{\alpha^{\prime}}^{\beta^{\prime}}Y_{I^{\prime}}^{J^{\prime}}%
\end{array}
\right)  ,
\]
} where $X_{\alpha}^{\beta}$ is an element of the first $M_{4}\left(
\mathbb{C}\right)  $ and $Y_{I^{\prime}}^{J^{\prime}}$ is an element of the
second $M_{4}\left(  \mathbb{C}\right)  .$ The reality operator $J$ is defined
by
\[
J=\left(
\begin{array}
[c]{cc}%
0 & \delta_{\alpha}^{\beta^{\prime}}\delta_{I}^{J^{\prime}}\\
\delta_{\alpha^{\prime}}^{\beta}\delta_{I^{\prime}}^{J} & 0
\end{array}
\right)  \times\text{\textrm{complex conjugation.}}%
\]
In this representation we deduce that $a^{o}$ takes the form {
\[
a^{o}=Ja^{\ast}J^{-1}=\left(
\begin{array}
[c]{cc}%
\delta_{\alpha}^{\beta}\widetilde{Y}_{I}^{J} & 0\\
0 & \widetilde{X}_{\alpha^{\prime}}^{\beta^{\prime}}\delta_{^{I^{\prime}%
\prime}}^{J^{\prime}}%
\end{array}
\right)  ,
\]
} where $\widetilde{}$ denotes transposition. It is trivial to verify that
$\left[  a,b^{o}\right]  =0.$

The order one condition is {
\[
\left[  \left[  D,a\right]  ,b^{o}\right]  =0
\]
} If we write {
\[
b^{o}=\left(
\begin{array}
[c]{cc}%
\delta_{\alpha}^{\beta}W_{I}^{J} & 0\\
0 & Z_{\alpha^{\prime}}^{\beta^{\prime}}\delta_{^{I^{\prime}}}^{J^{\prime}}%
\end{array}
\right)  ,
\]
} then {
\begin{align*}
\left[  \left[  D,a\right]  ,b^{o}\right]   &  =\left(
\begin{array}
[c]{cc}%
\left[  \left[  D,X\right]  ,W\right]  _{A}^{B} & \left(  \left(
DY-XD\right)  Z-W\left(  DY-XD\right)  \right)  _{A}^{B^{\prime}}\\
\left(  \left(  DX-YD\right)  W-Z\left(  DX-YD\right)  \right)  _{A^{\prime}%
}^{B} & \left[  \left[  D,Y\right]  ,Z\right]  _{A^{\prime}}^{B^{\prime}}%
\end{array}
\right) \\
&  =0.
\end{align*}
} Explicitely, the first two equations read: {
\begin{align*}
\left(  D_{\alpha I}^{\gamma K}X_{\gamma}^{\beta}-X_{\alpha}^{\gamma}D_{\gamma
I}^{\beta K}\right)  W_{K}^{J}-W_{I}^{K}\left(  D_{\alpha K}^{\gamma
J}X_{\gamma}^{\beta}-X_{\alpha}^{\gamma}D_{\gamma K}^{\beta J}\right)   &
=0\\
\left(  D_{\alpha I}^{\gamma^{\prime}K^{\prime}}Y_{K^{\prime}}^{J^{\prime}%
}-X_{\alpha}^{\gamma}D_{\gamma I}^{\gamma^{\prime}K}\right)  Z_{\gamma
^{\prime}}^{\beta^{\prime}}-W_{I}^{K}\left(  D_{\alpha K}^{\beta^{\prime
}K^{\prime}}Y_{K^{\prime}}^{J^{\prime}}-X_{\alpha}^{\gamma}D_{\gamma K}%
^{\beta^{\prime}J^{\prime}}\right)   &  =0.
\end{align*}
} We have shown \cite{beggar}, \cite{mc2}, that the only non-zero solution of
the second equation is {
\[
D_{\alpha I}^{\beta^{\prime}K^{\prime}}=\delta_{\alpha}^{\overset{.}{1}}%
\delta_{\overset{.}{1^{\prime}}}^{\beta^{\prime}}\delta_{I}^{1}\delta
_{1^{\prime}}^{K^{\prime}}k^{\ast\nu_{R}}%
\]
} which means that there can be only \emph{one non-zero single entry} in the
off-diagonal $16\times16$ matrix $D_{A}^{B^{^{\prime}}},$ and this implies
that {
\begin{align*}
D_{\alpha I}^{\beta J}  &  =D_{\alpha\left(  l\right)  }^{\beta}\delta_{I}%
^{1}\delta_{1}^{J}+D_{\alpha\left(  q\right)  }^{\beta}\delta_{I}^{i}%
\delta_{j}^{J}\delta_{i}^{j}\\
Y_{I^{\prime}}^{J^{\prime}}  &  =\delta_{I^{\prime}}^{1^{\prime}}%
\delta_{1^{\prime}}^{J^{\prime}}Y_{1^{\prime}}^{1^{\prime}}+\delta_{I^{\prime
}}^{i^{\prime}}\delta_{j^{\prime}}^{J^{\prime}}Y_{i^{\prime}}^{j^{\prime}}\\
X_{\overset{.}{1}}^{\overset{.}{1}}  &  =Y_{1^{\prime}}^{1^{\prime}},\hbox{
}X_{\overset{.}{1}}^{\alpha}=0,\quad\alpha\neq\overset{.}{1},
\end{align*}
where we have split the index }$I=1,i$, and $I^{\prime}=1^{\prime},i^{\prime
}.$ From the property of commutation of the grading operator {
\begin{align*}
g_{\alpha}^{\beta}  &  =\left(
\begin{array}
[c]{cc}%
1_{2} & 0\\
0 & -1_{2}%
\end{array}
\right) \\
\left[  g,a\right]   &  =0\quad a\in M_{4}\left(  \mathbb{C}\right)  ,
\end{align*}
} the algebra {$M_{4}\left(  \mathbb{C}\right)  $} reduces to { $M_{2}\left(
\mathbb{C}\right)  _{R}\oplus M_{2}\left(  \mathbb{C}\right)  _{L}.$} We
further impose the condition of symplectic isometry on {$M_{2}\left(
\mathbb{C}\right)  _{R}\oplus M_{2}\left(  \mathbb{C}\right)  _{L}$
\[
\sigma_{2}\otimes1_{2}\left(  \overline{a}\right)  \sigma_{2}\otimes1_{2}=a,
\]
} which reduces it to {$\mathbb{H}_{R}\mathbb{\oplus H}_{L}$}. We will be
using the notation {
\[
\alpha=\overset{.}{1},\overset{.}{2},a\hbox{ \ where }\xi_{\overset{.}%
{1},\overset{.}{2}}\text{\textrm{ }}\in{\mathbb{H}_{R},}\text{ }\xi_{a}%
\in\mathbb{H}_{L}.
\]
}Together with the above condition this implies that {
\[
X_{\alpha}^{\beta}=\delta_{\alpha}^{\overset{.}{1}}\delta_{\overset{.}{1}%
}^{\beta}X_{\overset{.}{1}}^{\overset{.}{1}}+\delta_{\alpha}^{\overset{.}{2}%
}\delta_{\overset{.}{2}}^{\beta}\overline{X}_{\overset{.}{1}}^{\overset{.}{1}%
}+\delta_{\alpha}^{a}\delta_{b}^{\beta}X_{a}^{b}%
\]
} and the algebra {$\mathbb{H}_{R}\mathbb{\oplus H}_{L}\oplus M_{4}\left(
\mathbb{C}\right)  $} reduces to {
\[
\mathbb{C\oplus H}\oplus M_{3}\left(  \mathbb{C}\right)
\]
} because {$X_{\overset{.}{1}}^{\overset{.}{1}}=Y_{1^{\prime}}^{1^{\prime}}.$
}Expanding the Dirac action we get{
\[
\psi_{A}^{\ast}D_{A}^{B}\psi_{B}+\psi_{\overset{.}{1^{\prime}}1^{\prime}%
}^{\ast}D_{\overset{.}{1^{\prime}}1^{\prime}}^{B}\psi_{B}+\psi_{A}^{\ast}%
D_{A}^{\overset{.}{1^{\prime}}1^{\prime}}\psi_{\overset{.}{1^{\prime}%
}1^{\prime}}+\psi_{A^{\prime}}^{\ast}D_{A^{\prime}}^{B^{\prime}}%
\psi_{B^{\prime}}%
\]
}The spinors can thus be denoted by {
\begin{align*}
\psi_{A}  &  =\psi_{\alpha I}=\left(  \psi_{\alpha1},\psi_{\alpha i}\right) \\
&  =\left(  \psi_{\overset{.}{1}1},\psi_{\overset{.}{2}1},\psi_{a1}%
,\psi_{\overset{.}{1}i},\psi_{\overset{.}{2}i},\psi_{ai}\right) \\
&  \equiv\left(  \nu_{R},e_{R},l_{a},u_{Ri},d_{Ri},q_{ai}\right)  ,
\end{align*}
} where $l_{a}=\left(  \nu_{L},e_{L}\right)  $ and $q_{ai}=\left(
u_{Li},d_{Li}\right)  .$ The component $\psi_{\overset{.}{1}^{\prime}%
1^{\prime}}=\psi_{\overset{.}{1}1}^{c}=\nu_{R}^{c}$ which implies that the
Dirac action {
\[
\psi_{A}^{\ast}D_{A}^{B}\psi_{B}+\nu_{R}^{\ast c}k^{\ast\nu_{R}}\nu
_{R}+\mathrm{c.c}%
\]
has only a mixing term for the right-handed neutrinos. }

Having determined the structure of the Dirac operator of the discrete space,
we can form the Dirac operator of the product space of this discrete space
times \ a four-dimensional Riemannian manifold: {
\[
D=D_{M}\otimes1+\gamma_{5}\otimes D_{F}.
\]
} Since $D_{F}$ is a $32\times32$ matrix tensored with the $3\times3$ matrices
of generation space and with the Clifford algebra, {$D$ is $384\times384$} matrix.

To take inner automorphisms into account, we have to evaluate the Dirac
operator {
\[
D_{A}=D+A+JAJ^{-1},
\]
} where
\[
A=%
{\displaystyle\sum}
a\left[  D,b\right]  .
\]
{In particular} {
\[
A_{A}^{B}=%
{\displaystyle\sum}
a_{A}^{C}\left(  D_{C}^{D}b_{D}^{B}-b_{C}^{D}D_{D}^{B}\right)  .
\]
} Note there are no mixing terms like $D_{C}^{D^{\prime}}b_{D^{\prime}}^{B}$
because $b$ is block diagonal.

Evaluating all components of the full Dirac operator $D_{M}^{N}$ , quoting
only the result, the full derivation will be given in a forthcoming paper
\cite{Prepare}, we obtain: { }%
\begin{align*}
\left(  D\right)  _{\overset{.}{1}1}^{\overset{.}{1}1}  &  =\gamma^{\mu
}\otimes D_{\mu}\otimes1_{3},\quad D_{\mu}=\partial_{\mu}+\frac{1}{4}%
\omega_{\mu}^{cd}\left(  e\right)  \gamma_{cd},\quad1_{3}=\text{generations}\\
\left(  D\right)  _{\overset{.}{1}1}^{a1}  &  =\gamma_{5}\otimes k^{\ast\nu
}\otimes\epsilon^{ab}H_{b}\qquad k^{\nu}=3\times3\text{ neutrino mixing
matrix}\\
\left(  D\right)  _{\overset{.}{2}1}^{\overset{.}{2}1}  &  =\gamma^{\mu
}\otimes\left(  D_{\mu}+ig_{1}B_{\mu}\right)  \otimes1_{3}\\
\left(  D\right)  _{\overset{.}{2}1}^{a1}  &  =\gamma_{5}\otimes k^{\ast
e}\otimes\overline{H}^{a}\\
\left(  D\right)  _{a1}^{\overset{.}{1}1}  &  =\gamma_{5}\otimes k^{\nu
}\otimes\epsilon_{ab}\overline{H}^{b}\\
\left(  D\right)  _{a1}^{\overset{.}{2}1}  &  =\gamma_{5}\otimes k^{e}\otimes
H_{a}\\
\left(  D\right)  _{a1}^{b1}  &  =\gamma^{\mu}\otimes\left(  \left(  D_{\mu
}+\frac{i}{2}g_{1}B_{\mu}\right)  \delta_{a}^{b}-\frac{i}{2}g_{2}W_{\mu
}^{\alpha}\left(  \sigma^{\alpha}\right)  _{a}^{b}\right)  \otimes1_{3},\text{
\qquad}\sigma^{\alpha}=\text{Pauli}\\
\left(  D\right)  _{\overset{.}{1}i}^{\overset{.}{1}j}  &  =\gamma^{\mu
}\otimes\left(  \left(  D_{\mu}-\frac{2i}{3}g_{1}B_{\mu}\right)  \delta
_{i}^{j}-\frac{i}{2}g_{3}V_{\mu}^{m}\left(  \lambda^{m}\right)  _{i}%
^{j}\right)  \otimes1_{3},\qquad\lambda^{i}=\text{Gell-Mann}\\
\left(  D\right)  _{\overset{.}{1}i}^{aj}  &  =\gamma_{5}\otimes k^{\ast
u}\otimes\epsilon^{ab}H_{b}\delta_{i}^{j}\\
\left(  D\right)  _{\overset{.}{2}i}^{\overset{.}{2}j}  &  =\gamma^{\mu
}\otimes\left(  \left(  D_{\mu}+\frac{i}{3}g_{1}B_{\mu}\right)  \delta_{i}%
^{j}-\frac{i}{2}g_{3}V_{\mu}^{m}\left(  \lambda^{m}\right)  _{i}^{j}\right)
\otimes1_{3}\\
\left(  D\right)  _{\overset{.}{2}i}^{aj}  &  =\gamma_{5}\otimes k^{\ast
d}\otimes\overline{H}^{a}\delta_{i}^{j}\\
\left(  D\right)  _{ai}^{bj}  &  =\gamma^{\mu}\otimes\left(  \left(  D_{\mu
}-\frac{i}{6}g_{1}B_{\mu}\right)  \delta_{a}^{b}\delta_{i}^{j}-\frac{i}%
{2}g_{2}W_{\mu}^{\alpha}\left(  \sigma^{\alpha}\right)  _{a}^{b}\delta_{i}%
^{j}-\frac{i}{2}g_{3}V_{\mu}^{m}\left(  \lambda^{m}\right)  _{i}^{j}\delta
_{a}^{b}\right)  \otimes1_{3}\\
\left(  D\right)  _{ai}^{\overset{.}{1}j}  &  =\gamma_{5}\otimes k^{u}%
\otimes\epsilon_{ab}\overline{H}^{b}\delta_{i}^{j}\\
\left(  D\right)  _{ai}^{\overset{.}{2}j}  &  =\gamma_{5}\otimes k^{d}\otimes
H_{a}\delta_{i}^{j}\\
\left(  D\right)  _{\overset{.}{1}1}^{\overset{.}{1^{\prime}}1^{\prime}}  &
=\gamma_{5}\otimes k^{\ast\nu_{R}}\sigma\qquad\text{generate scale }%
M_{R}\text{ by }\sigma\rightarrow M_{R}\\
\left(  D\right)  _{\overset{.}{1^{\prime}}1^{\prime}}^{\overset{.}{1}1}  &
=\gamma_{5}\otimes k^{\nu_{R}}\sigma\\
D_{A^{\prime}}^{B^{\prime}}  &  =\overline{D}_{A}^{B},\qquad D_{A^{\prime}%
}^{B}=\overline{D}_{A}^{B^{\prime}},\qquad D_{A}^{B^{\prime}}=\overline
{D}_{A^{\prime}}^{B}%
\end{align*}
{where }$B_{\mu},W_{\mu}^{\alpha}$ and $V_{\mu}^{m}$ are the $U(1),$ $SU(2)$
and $SU(3)$ gauge fields, and $H$ is a complex doublet scalar field and
$\sigma$ is a singlet real scalar field. We have assumed that the unitary
algebra $\mathcal{U}\left(  \mathcal{A}\right)  $ is restricted to
$\mathcal{SU}\left(  \mathcal{A}\right)  $ to eliminate a superfluous $U(1)$
gauge field. Pictorially, {the matrix }$D_{M}^{N}$ has the structure:{ }{
\begin{align*}
&  \qquad\qquad\quad\left(
\begin{array}
[c]{cccccc}%
\begin{array}
[c]{c}%
\overset{.}{1}1\\
v_{R}%
\end{array}
&
\begin{array}
[c]{c}%
\overset{.}{2}1\\
e_{R}%
\end{array}
&
\begin{array}
[c]{c}%
a1\\
l_{a}%
\end{array}
&
\begin{array}
[c]{c}%
\overset{.}{1}i\\
u_{iR}%
\end{array}
&
\begin{array}
[c]{c}%
\overset{.}{2}i\\
d_{iR}%
\end{array}
&
\begin{array}
[c]{c}%
ai\\
q_{iL}%
\end{array}
\end{array}
\right) \\
&  \left(
\begin{array}
[c]{c}%
\overset{.}{1}1\\
\overset{.}{2}1\\
b1\\
\overset{.}{1}j\\
\overset{.}{2}j\\
bj
\end{array}
\right)  \left(
\begin{array}
[c]{cccccc}%
\left(  D\right)  _{\overset{.}{1}1}^{\overset{.}{1}1} & 0 & \left(  D\right)
_{\overset{.}{1}1}^{a1} & 0 & 0 & 0\\
0 & \left(  D\right)  _{\overset{.}{2}1}^{\overset{.}{2}1} & \left(  D\right)
_{\overset{.}{2}1}^{a1} & 0 & 0 & 0\\
\left(  D\right)  _{b1}^{\overset{.}{1}1} & \left(  D\right)  _{b1}%
^{\overset{.}{2}1} & \left(  D\right)  _{a1}^{b1} & 0 & 0 & 0\\
0 & 0 & 0 & \left(  D\right)  _{\overset{.}{1}j}^{\overset{.}{1}i} & 0 &
\left(  D\right)  _{\overset{.}{1}j}^{ai}\\
0 & 0 & 0 & 0 & \left(  D\right)  _{\overset{.}{2}j}^{\overset{.}{2}i} &
\left(  D\right)  _{\overset{.}{2}j}^{ai}\\
0 & 0 & 0 & \left(  D\right)  _{bj}^{\overset{.}{1}i} & \left(  D\right)
_{bj}^{\overset{.}{2}i} & \left(  D\right)  _{bj}^{ai}%
\end{array}
\right)
\end{align*}
}

Needless to say the term $\psi_{M}^{\ast}D_{M}^{N}\psi_{N}$ contains all the
fermionic terms and their interactions in the standard model.

\section{{\emph{The Spectral Action Principle}}}

There is a shift of point of view in NCG similar to Fourier transform, where
the usual emphasis on the points {$x\in M$ of a geometric space is now
replaced by the spectrum $\Sigma$ of the operator $D.$} The existence of
Riemannian manifolds which are isospectral but not isometric shows that the
following hypothesis is stronger than the usual diffeomorphism invariance of
the action of general relativity

\begin{center}
\bigskip\emph{The physical action depends only on the }$\Sigma$

\bigskip
\end{center}

This is the {spectral action principle \cite{cc2}}. The spectrum is a
geometric invariant and replaces {diffeomorphism invariance}. We now apply
this basic principle to the noncommutative geometry defined by the spectrum of
the standard model to show that the dynamics of all interactions, including
gravity is given by the spectral action%
\[
{\hbox{Trace }f\left(  \frac{D_{A}}{\Lambda}\right)  +\frac{1}{2}\left\langle
J\Psi,D_{A}\Psi\right\rangle ,}%
\]
where {$f$} is a positive function, {$\Lambda$} a cutoff scale needed to make
$\frac{D_{A}}{\Lambda}$ dimensionless, and ${\Psi}$ is a Grassmann variable
which represents fermions.

In the case of the cut-off function, $f$ only plays a role through its momenta
{$f_{0},f_{2},f_{4}$} where
\[
{f_{k}=%
{\displaystyle\int\limits_{0}^{\infty}}
f(v)v^{k-1}dv,\quad\hbox{for }k>0,\quad\hbox{, \ }f_{0}=f(0).}%
\]
These will serve as three free parameters in the model. {\ In this case the
action $S_{\Lambda}[D_{A}]$} is the number of eigenvalues {$\lambda$ of
$D_{A}$} counted with their multiplicities such that {$|\lambda|\leq\Lambda$.}

\bigskip To illustrate how this comes about, expand the function $f$ in terms
of its Laplace transform{%
\begin{align*}
{\hbox{Trace}}f\left(  P\right)   &  =%
{\displaystyle\sum\limits_{s}}
f_{s^{\prime}}{\hbox{Trace}}\left(  P^{-s}\right) \\
{\hbox{Trace}}\left(  P^{-s}\right)   &  =\frac{1}{\Gamma\left(  s\right)  }%
{\displaystyle\int\limits_{0}^{\infty}}
t^{s-1}{\hbox{Trace}}\left(  e^{-tP}\right)  dt\qquad\hbox{Re}\left(
s\right)  \geq0\\
{\hbox{Trace}}\left(  e^{-tP}\right)   &  \simeq%
{\displaystyle\sum\limits_{n\geq0}}
t^{\frac{n-m}{d}}%
{\displaystyle\int\limits_{M}}
a_{n}\left(  x,P\right)  dv\left(  x\right)  ,
\end{align*}
} where $m=4$ is the dimension of the manifold $M$ and $d=2$ is the order of
the elliptic operator $D^{2}.$ Gilkey gives generic formulas for the
Seeley-deWitt coefficients $a_{n}\left(  x,P\right)  $ for a large class \ of
differential operators $P$ \cite{Gilkey}. The details are explained in
preceding papers \cite{cc2}, \cite{mc2} or using the tensorial notation, in a
forthcoming paper \cite{Prepare}.

The bosonic part \ of the spectral action, gives an action that unifies
gravity with {$SU(2)\times U(1)\times SU(3)$} Yang-Mills gauge theory, with a
{Higgs doublet }$H$ and spontaneous symmetry breaking and a real scaler field
$\sigma.$ It is given by \cite{cc2}, \cite{mc2}
\begin{align*}
S  &  =\frac{48}{\pi^{2}}f_{4}\Lambda^{4}%
{\displaystyle\int}
d^{4}x\sqrt{g}\\
&  -\frac{4}{\pi^{2}}f_{2}\Lambda^{2}%
{\displaystyle\int}
d^{4}x\sqrt{g}\left(  R+\frac{1}{2}a\overline{H}H+\frac{1}{4}c\right) \\
&  +\frac{1}{2\pi^{2}}f_{0}%
{\displaystyle\int}
d^{4}x\sqrt{g}\left[  \frac{1}{30}\left(  -18C_{\mu\nu\rho\sigma}%
^{2}+11R^{\ast}R^{\ast}\right)  +\frac{5}{3}g_{1}^{2}B_{\mu\nu}^{2}+g_{2}%
^{2}\left(  W_{\mu\nu}^{\alpha}\right)  ^{2}+g_{3}^{2}\left(  V_{\mu\nu}%
^{m}\right)  ^{2}\right. \\
&  \qquad\left.  +\frac{1}{6}aRH_{a}\overline{H}^{a}+b\left(  \overline
{H}H\right)  ^{2}+a\left\vert \nabla_{\mu}H_{a}\right\vert ^{2}+2e\overline
{H}H\,\sigma^{2}+\frac{1}{2}d\,\sigma^{4}+\frac{1}{12}cR\sigma^{2}+\frac{1}%
{2}c\left(  \partial_{\mu}\sigma\right)  ^{2}\right] \\
&  +f_{-2}\Lambda^{-2}a_{6}+\cdots
\end{align*}
This can be rearranged, after normalizing the kinetic energies and ignoring
the $\sigma$ field which only plays a role in cosmology, to the form:
\[%
\begin{array}
[c]{rl}%
S= & \int\ \biggl(\frac{1}{2\kappa_{0}^{2}}\,R+\alpha_{0}\,C_{\mu\nu\rho
\sigma}\,C^{\mu\nu\rho\sigma}+\ \ \gamma_{0}+\tau_{0}\,R^{\ast}R^{\ast}\\[3mm]%
+ & \frac{1}{4}\,G_{\mu\nu}^{i}\,G^{\mu\nu i}+\frac{1}{4}\,F_{\mu\nu}^{\alpha
}\,F^{\mu\nu\alpha}+\ \frac{1}{4}\,B_{\mu\nu}\,B^{\mu\nu}\\[2mm]%
+ & \,\frac{1}{2}|D_{\mu}\,\mathbf{H}|^{2}-\mu_{0}^{2}|\mathbf{H}|^{2}-\xi
_{0}\,R\,|\mathbf{H}|^{2}+\lambda_{0}|\mathbf{H}|^{4}\biggl)\sqrt{g}\,d^{4}x,
\end{array}
\]
where%
\begin{align*}
\frac{1}{\kappa_{0}^{2}}  &  =\Lambda^{2}\ \frac{96\,f_{2}\,-f_{0}\,c}%
{12\,\pi^{2}},\qquad\mu_{0}^{2}=\Lambda^{2}\ \left(  2\,\frac{f_{2}\,}{f_{0}%
}-\,\frac{e}{a}\right) \\
\alpha_{0}  &  =-\frac{3\,f_{0}}{10\,\pi^{2}},\qquad\tau_{0}=\frac{11\,f_{0}%
}{60\,\pi^{2}},\qquad\lambda_{0}=\frac{\pi^{2}}{2\,f_{0}}\frac{b}{a^{2}}\\
\gamma_{0}  &  =\ \Lambda^{4}\frac{1}{\pi^{2}}(48\,f_{4}\,-f_{2}\,\,c+\frac
{1}{4}f_{0}d),\qquad\xi_{0}=\frac{1}{12}.
\end{align*}
The parameters {$a,$ $b,$ $c,$ $d,$ $e$} are all dimensionless and related to
the Yukawa couplings that give the fermionic masses after the spontaneous
breaking of symmetry:{%
\begin{align*}
a  &  =\hbox{Tr}\left(  k^{e\ast}k^{e}+k^{\nu\ast}k^{\nu}+3k^{u\ast}%
k^{u}+3k^{d\ast}k^{d}\right) \\
b  &  =\hbox{Tr}\left(  \left(  k^{e\ast}k^{e}\right)  ^{2}+\left(  k^{\nu
\ast}k^{\nu}\right)  ^{2}+3\left(  k^{u\ast}k^{u}\right)  ^{2}+3\left(
k^{d\ast}k^{d}\right)  ^{2}\right) \\
c  &  =\hbox{Tr}\left(  k_{R}^{\ast}k_{R}\right)  ,\quad d=\hbox{Tr}\left(
(k_{R}^{\ast}k_{R}\right)  ^{2}),\quad e=\hbox{Tr}\left(  k_{R}^{\ast}%
k_{R}k^{\nu\ast}k^{\nu}\right)  .
\end{align*}
}

\section{{\emph{Predictions of Spectral Action for Standard Model}}}

We shall first perform our analysis by assuming that the function $f$ \ is
well approximated by the cut-off function, thus allowing us to ignore higher
order terms. We will determine, to what extent such an approximation could be
made, and its effects on the predictions. The normalization of the kinetic
terms imposes a relation between the coupling constants {$g_{1}$, $g_{2}$,
$g_{3}$} and the coefficient $f_{0}$, of the form
\[
{\frac{g_{3}^{2}\,f_{0}}{2\pi^{2}}=\frac{1}{4},\ \ \ \ \ \ g_{3}^{2}=g_{2}%
^{2}=\frac{5}{3}\,g_{1}^{2}\,.}%
\]
This gives the relation {$\sin^{2}\theta_{W}=\frac{3}{8}$} a value also
obtained in {$SU(5)$ }and {$SO(10)$} grand unified theories. The three momenta
of the function {$f_{0},$ $f_{2}$} and {$f_{4}$} can be used to specify the
initial conditions on the gauge couplings, the {Newton constant} and the
{cosmological constant}. We deduce that the geometrical picture is valid at
high energies, and the spectral action must be considered in the {Wilsonian}
approach, where all coupling constants are energy dependent and follow the
{renormalization group equations.} \ For example, The fine structure constant
{$\alpha_{em}$ }is given by
\[
{\alpha_{em}=\,\sin(\theta_{w})^{2}\,\alpha_{2}\,,\quad\alpha_{i}=\frac
{g_{i}^{2}}{4\pi}.}%
\]
Its infrared value is {$\sim1/137.036$} but it is running as a function of the
energy and increases to the value {$\alpha_{em}(M_{Z})=1/128.09$ }already, at
the energy {$M_{Z}\sim91.188$ Gev.}

Assuming the {\textquotedblleft big desert\textquotedblright} hypothesis, the
running of the three couplings $\alpha_{i}$ is known. With 1-loop corrections
only, it is given by \cite{Ross}
\[
{\beta_{g_{i}}=(4\pi)^{-2}\,b_{i}\,g_{i}^{3},\ \ \ {\hbox{ with }}%
\ \ b=(\frac{41}{6},-\frac{19}{6},-7),}%
\]
so that {%
\begin{align*}
\alpha_{1}^{-1}(\Lambda)  &  =\,\alpha_{1}^{-1}(M_{Z})-\frac{41}{12\pi}%
\,\log\,\frac{\Lambda}{M_{Z}}\\
\alpha_{2}^{-1}(\Lambda)  &  =\,\alpha_{2}^{-1}(M_{Z})+\frac{19}{12\pi}%
\,\log\,\frac{\Lambda}{M_{Z}}\\
\alpha_{3}^{-1}(\Lambda)  &  =\,\alpha_{3}^{-1}(M_{Z})+\frac{42}{12\pi}%
\,\log\,\frac{\Lambda}{M_{Z}},
\end{align*}
} where $M_{Z}$ is the mass of the $Z^{0}$ vector boson.

\smallskip\ In fact, if one considers the actual experimental values {%
\[
g_{1}(M_{Z})=0.3575,\ \ \ g_{2}(M_{Z})=0.6514,\ \ \ g_{3}(M_{Z})=1.221,
\]
} one obtains the values {%
\[
\alpha_{1}(M_{Z})=0.0101,\ \ \ \alpha_{2}(M_{Z})=0.0337,\ \ \ \alpha_{3}%
(M_{Z})=0.1186.
\]
} The graphs of the running of the three constants $\alpha_{i}$ do not meet
exactly, hence do not specify a unique unification energy.%

\begin{figure}
[ptb]
\begin{center}
\includegraphics[
height=2.2675in,
width=3.6564in
]%
{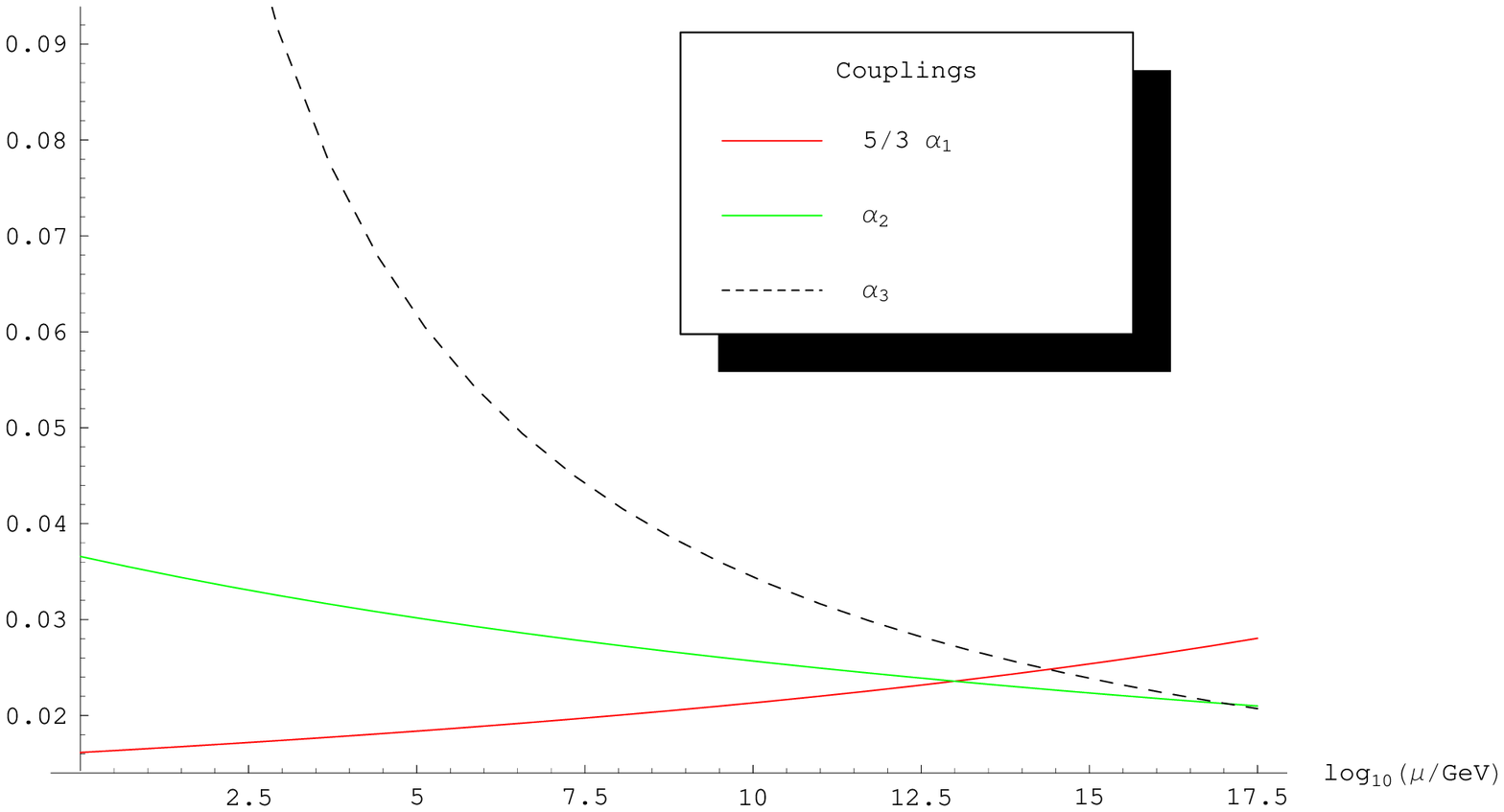}%
\end{center}
\end{figure}

Next we study the running of the Higgs quartic coupling $\lambda$ \cite{Sher}:
{%
\[
\frac{d\lambda}{dt}=\lambda\gamma+\frac{1}{8\pi^{2}}(12\lambda^{2}+B),
\]
} where{%
\begin{align*}
\gamma &  =\,\frac{1}{16\pi^{2}}(12k_{t}^{2}-9g_{2}^{2}-3g_{1}^{2})\\
B  &  =\,\frac{3}{16}(3g_{2}^{4}+2g_{1}^{2}\,g_{2}^{2}+g_{1}^{4})-3\,k_{t}%
^{4}\,.
\end{align*}
}

The Higgs mass is then given by {%
\[
m_{H}^{2}=\,8\lambda\,\frac{M^{2}}{g^{2}}\,,\quad m_{H}=\sqrt{2\lambda}%
\,\frac{2M}{g}.
\]
}The numerical solution to these equations with the boundary value
{$\lambda_{0}=0.356$} at {$\Lambda=10^{17}$ Gev }gives {$\lambda(M_{Z}%
)\sim0.241$} and a Higgs mass of the order of {$170$ Gev. This specific value
has been recently ruled out experimentally. However, this is to be expected,
because of the non unification of the three gauge couplings. }

The mass of the top quark is governed by the top quark Yukawa coupling $k_{t}$
and is given by the equation {%
\[
m_{top}(t)=\frac{1}{\sqrt{2}}\frac{2M}{g}\,k_{t}=\frac{1}{\sqrt{2}}%
\,v\,k_{t},
\]
} where {$v=\frac{2M}{g}$ }is the vacuum expectation value of the Higgs field.
\ There is a relation between the Yukawa and the gauge couplings which emerges
as a consequence of normalizing the Higgs interactions. This relation is a
consequence of the fact that all fermions get their masses by coupling to the
same Higgs through interactions of the form {
\[
kH\overline{\psi}\psi.
\]
} After normalizing the kinetic energies of the Higgs field through the
redefinition $H\rightarrow${$\frac{\pi}{\sqrt{af_{0}}}H,$} the mass terms take
the form {
\[
\frac{\pi}{\sqrt{f_{0}}}\frac{k}{\sqrt{a}}H\overline{\psi}\psi.
\]
} Using the identity $%
{\displaystyle\sum\limits_{i}}
\left(  \frac{k_{i}}{\sqrt{a}}\right)  ^{2}=1$ gives a relation among the
fermions masses and W mass \cite{mc2}
\[
{\sum_{\mathrm{generations}}m_{e}^{2}+m_{\nu}^{2}+3m_{d}^{2}+3m_{u}^{2}%
=8M_{W}^{2}.}%
\]
The value of $g$ at a unification scale of {$10^{17}$ Gev} is {$\sim0.517$.}
Thus, neglecting the $\tau$ neutrino Yukawa coupling, we get the simplified
relation {%
\[
k_{t}=\,\frac{2}{\sqrt{3}}\,g\sim0.597\,\,.
\]
} The numerical integration of the differential equation with the boundary
condition gives the value $k${$_{0}\sim1.102$} and a top quark mass of the
order of {$\frac{1}{\sqrt{2}}k_{0}\,v\sim\,173.683\,k_{0}$ Gev. The value of
}$k_{0}$ improves to $k${$_{0}\sim1.04$ when the }$\tau$-neutrino Yukawa
coupling is taken into account, which yields an acceptable value for the top
quark mass of {$179$ Gev \cite{mc2}. One reason why the resulting top quark
mass is acceptable while the Higgs mass is not, is because the later is
dependent on the cut-off function. }

The fact that the coupling constants do not meet is giving us information
about the nature of the function $f$ used in the spectral action. Our results
were obtained under the assumption that the function $f$ is the cut-off
function for which all coefficients of the higher order terms in the
asymptotic expansion vanish. These coefficients are given by derivatives of
the function evaluated at zero. We can infer from these results, especially
from the near meeting of the coupling constants, the good approximate values
for $\sin^{2}\theta$ and the top quark mass, that the function $f$ is well
approximated by the cut-off function, but deviates slightly from it. What is
needed then is for the Taylor coefficients of the function to be very small
but not zero.

To prove that this is indeed the case we compute the gauge and Higgs
contributions to the next order i.e. $a_{6},$ in the asymptotic expansion. It
is enough to look only at the non gravitational terms \cite{Prepare}: {
\begin{align*}
&  -\frac{f^{\prime}(0)}{12\pi^{2}\Lambda^{2}}\left[  c_{1}\overline
{H}H\left(  \frac{1}{4}g_{2}^{2}\left(  W_{\mu\nu}^{\alpha}\right)
^{2}\right)  +c_{2}\overline{H}H\left(  g_{3}^{2}\left(  V_{\mu\nu}%
^{m}\right)  ^{2}\right)  +c_{3}\overline{H}\sigma^{\alpha}H\left(  \frac
{1}{2}g_{1}g_{2}B_{\mu\nu}W_{\mu\nu}^{\alpha}\right)  \right. \\
&  \qquad\qquad+c_{4}\left(  \overline{H}H\right)  ^{3}+c_{5}\left(
\overline{H}H\right)  ^{2}\sigma^{2}+c_{6}\left(  \left(  \overline{H}%
\nabla_{\mu}H\right)  ^{2}+\left(  \nabla_{\mu}\overline{H}H\right)
^{2}\right) \\
&  \qquad\qquad+c_{7}\left(  \nabla_{\mu}\nabla_{\nu}\overline{H}\right)
\left(  \nabla_{\mu}\nabla_{\nu}H\right)  +c_{8}\left(  \overline
{H}H\left\vert \nabla_{\mu}H\right\vert ^{2}+\left\vert \overline{H}%
\nabla_{\mu}H\right\vert ^{2}\right)  +c_{9}\left\vert \nabla_{\mu}\left(
H\sigma\right)  \right\vert ^{2}\\
&  \qquad\qquad\left.  +c_{10}\left\vert \epsilon^{ab}H_{a}\nabla_{\mu}%
H_{b}\right\vert ^{2}+c_{11}\nabla_{\mu}\overline{H}\nabla_{\nu}H\left(
\frac{3}{2}ig_{1}B_{\mu\nu}\right)  +c_{12}\nabla_{\mu}\overline{H}%
\sigma^{\alpha}\nabla_{\nu}H\left(  \frac{3}{2}ig_{2}W_{\mu\nu}^{\alpha
}\right)  \right]
\end{align*}
} where the coefficients $c_{1},\cdots,c_{12}$ depend only on the Yukawa
couplings. The exact expression will be given in reference \cite{Prepare}.
This clearly shows that the kinetic terms of the gauge fields get modified,
and are all multiplied with the coefficients {$f_{-2}=f^{\prime}(0).$} The
remarkable thing is that if we rescale the Higgs field by{
\[
H=\varphi\frac{\Lambda}{\left\vert k^{t}\right\vert },
\]
} assuming the top quark mass dominate the other fermion masses, then the
potential will depend on $\Lambda$ through an overall scale and the {
$\left\vert k^{t}\right\vert $ dependence drops out%
\[
V=\frac{3\Lambda^{4}}{\pi^{2}}\left(  -2f_{2}\overline{\varphi}\varphi
+\frac{1}{2}f_{0}\left(  \overline{\varphi}\varphi\right)  ^{2}+\frac{1}%
{3}f_{-2}\left(  \overline{\varphi}\varphi\right)  ^{3}+\cdots\right)  .
\]
} Now since $\varphi$ is a dimensionless doublet field, the vev
\[
\left\langle \varphi\right\rangle =v\left(
\begin{array}
[c]{c}%
0\\
1
\end{array}
\right)  ,
\]
will have a numerical value that depends only on the coefficients $f_{2},$
$f_{0}$\
\[
v_{0}^{2}=\frac{f_{0}}{2f_{-2}},
\]
and will be perturbed very slightly by the higher coefficients $f_{-2}$,
$f_{-4}\cdots$, provided they decrease very rapidly. Looking at the minimum of
the potential with the three terms above we have
\[
v^{2}=\frac{f_{0}}{2f_{-2}}\left(  -1+\sqrt{1+8\frac{f_{2}f_{-2}}{f_{0}^{2}}%
}\right)  .
\]
Thus the condition that the higher order term in the potential perturb the
minimum $v_{0}$ slightly requires the condition
\[
f_{-2}\ll\frac{f_{0}^{2}}{8f_{2}},
\]
so that
\[
v^{2}\simeq v_{0}^{2}\left(  1-4\frac{f_{2}f_{-2}}{f_{0}^{2}}\right)  .
\]
We can get a rough estimate of the coefficients $f_{0}$ and $f_{2}$ at
unification scale by setting
\[
\frac{4f_{2}\Lambda^{2}}{\pi^{2}}=\frac{1}{2\kappa^{2}},\qquad\kappa
=4.2\times10^{-19}\text{Gev}^{-1}%
\]
\ which implies that
\[
f_{2}\simeq\left(  \frac{\pi^{2}}{8}\right)  \left(  \frac{1}{\kappa\Lambda
}\right)  ^{2}.
\]
Thus if $\Lambda$ is of the order of $M_{\text{Planck}}$ then $f_{2}\sim1$
while if $\Lambda\sim10^{17}$ then $f_{2}\sim10^{2}.$ We also have
\[
\frac{f_{0}g_{3}^{2}}{2\pi^{2}}=\frac{1}{4},
\]
thus
\[
f_{0}=\frac{\pi}{8\alpha_{s}}\sim20,\qquad\alpha_{s}=\frac{g_{3}^{2}}{4\pi}%
\]
at unification scale. Therefore we must have
\[
f_{-2}\ll\frac{10^{2}}{f_{2}}%
\]
and this can be anywhere between $10^{2}$ and $10^{-2}$ depending whether
$\Lambda$ is at the Planck mass or two orders less.

We can now speculate on the form of the function $F(D^{2})=f\left(  D\right)
$. This function must have rapidly decreasing Taylor coefficients (these are
$F_{0}=F\left(  0\right)  ,$ $F_{-2}=-F^{\prime}\left(  0\right)  ,$
$F_{-4}=F^{\prime\prime}\left(  0\right)  \cdots$ ) while the Mellin
coefficients $F_{2},$ $F_{4}$ should behave independently. The cut-off
function can be approximated by the sequence $F_{\left\{  N\right\}  }\left(
x\right)  $
\[
F_{\left\{  N\right\}  }\left(  x\right)  =A\left(  1+x+\frac{1}{2!}%
x^{2}+\cdots+\frac{1}{N!}x^{N}\right)  e^{-x}%
\]
where
\[
A\sim20.
\]
This function has the property that the first $N$ coefficients in the Taylor
expansion vanish, and is thus a very good approximation to a cut-off function.
A slightly perturbed form of this function is given by
\[
F_{\left\{  N\right\}  }\left(  x,\epsilon\right)  =e^{-\epsilon x}F_{\left\{
N\right\}  }\left(  x\right)
\]
where $\epsilon\leq\pm10^{-2}$. In this case, we have $f_{-2}=A\epsilon,$
$f_{-4}=A\epsilon^{2}.$ To have a feeling about this function we can plot
$F_{\left\{  10\right\}  }\left(  x,\epsilon\right)  $

{\parbox[b]{4.4996in}{\begin{center}
\fbox{\includegraphics[
height=3in,
width=4.4996in
]%
{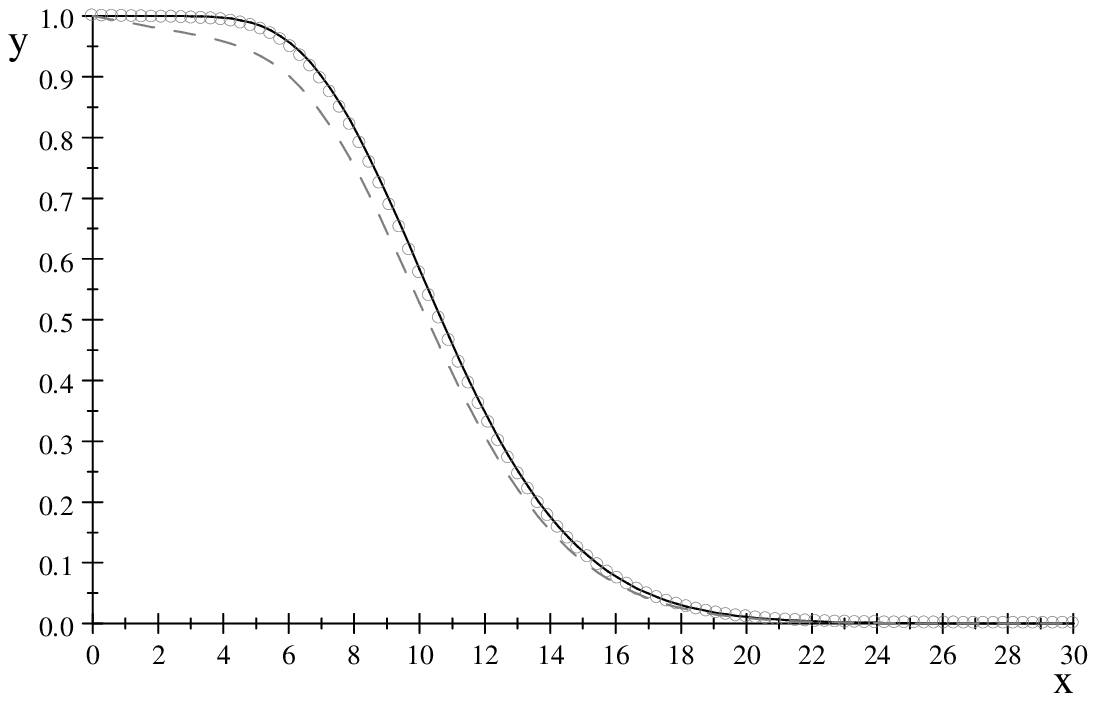}%
}\\
$F_{\left\{  10\right\}  }\left(  x,\epsilon\right)  $ $\epsilon=0$ (solid)
$\epsilon=0.01$ (dash) $\epsilon=0.001$ (circles)
\end{center}}}%

\bigskip This shows that $\epsilon$ should be at least of order$10^{-2}$ to
$10^{-3}$ in order not to disturb the cut-off function much, in the region
where the scale is comparable to $\Lambda.$ As seen from the plot, the
function $F_{N}\left(  x,\epsilon\right)  $ is indistinguishable from
$F_{N}\left(  x\right)  $ for $\epsilon\sim10^{-3}.$ From this we deduce that
higher order terms in the heat kernel expansion will be supressed by the
Taylor coefficients of the function, and the perturbation can be trusted to
within one order from the Planck scale. This property will insure that the
initial conditions on the RG equations for the gauge coupling constant get
modified. To see this, we have, to lowest order, the modification to the gauge
kinetic energies \cite{Prepare}: {
\begin{align*}
&  \frac{f_{-2}v_{0}^{2}}{12\pi^{2}}\left[  \left(  \frac{1}{4}g_{1}^{2}%
B_{\mu\nu}^{2}\right)  \left(  \frac{17}{3}\right)  +\left(  \frac{1}{4}%
g_{2}^{2}\left(  W_{\mu\nu}^{\alpha}\right)  ^{2}\right)  \left(  3\right)
+\left(  \frac{1}{4}g_{3}^{2}\left(  V_{\mu\nu}^{m}\right)  ^{2}\right)
\left(  4\right)  \right.  \\
&  +\frac{1}{2}g_{1}g_{2}B_{\mu\nu}W_{\mu\nu}^{3}-\frac{3}{2}v^{2}\left(
g_{1}B_{\mu}-g_{2}W_{\mu}^{3}\right)  ^{2}+6g_{2}^{2}W_{\mu}^{+}W_{\mu}%
^{-}+6v^{2}\left(  g_{1}B_{\mu}-g_{2}W_{\mu}^{3}\right)  ^{2}\\
&  +\frac{9i}{4}g_{1}g_{2}^{2}B_{\mu\nu}W_{\mu}^{+}W_{\nu}^{-}+\frac{3}%
{2}g_{2}^{2}\left\vert \partial_{\mu}W_{\nu}^{-}-\frac{i}{2}\left(
g_{1}B_{\mu}-g_{2}W_{\mu}^{3}\right)  W_{\nu}^{-}-\frac{i}{2}W_{\mu}%
^{-}\left(  g_{1}B_{\nu}-g_{2}W_{\nu}^{3}\right)  \right\vert ^{2}\\
&  \left.  +\frac{3}{4}\left\vert \partial_{\mu}\left(  g_{1}B_{\nu}%
-g_{2}W_{\nu}^{3}\right)  -ig_{2}^{2}W_{\mu}^{+}W_{\nu}^{-}-\frac{i}{2}\left(
g_{1}B_{\mu}-g_{2}W_{\mu}^{3}\right)  \left(  g_{1}B_{\nu}-g_{2}W_{\nu}%
^{3}\right)  \right\vert ^{2}\right]
\end{align*}
It remains to show that this form, for some value of }$f_{-2},$ can provide a
mechanism for the unification of the three gauge couplings at some energy not
far from the Planck scale. Similarly, the contributions to the Higgs potential
are expected to modify the prediction of the Higgs mass \cite{sturmia}. The
analysis of the running of the gauge coupling constants and the Higgs mass,
taking these higher order terms into account is presently under study. We hope
to report on this in the near future.

\section{{ Spectral Action for Noncommutative Spaces with Boundary}}

In the {Hamiltonian quantization} of gravity it is essential to include
{boundary terms} in the action as this allows to define consistently the
momentum conjugate to the metric. This makes it necessary to modify the
{Einstein-Hilbert} action by adding to it a surface integral term so that the
variation of the action is well defined. The reason for this is that the
curvature scalar {$R$} contains second derivatives of the metric, which are
removed after integrating by parts to obtain an action which is quadratic in
first derivatives of the metric. To see this note that the curvature
{$R\sim\partial\Gamma+\Gamma\Gamma$} where {$\Gamma\sim g^{-1}\partial g$} has
two parts, one part is of second order in derivatives of the form
{$g^{-1}\partial^{2}g$} and the second part is the square of derivative terms
of the form {$\partial g\partial g.$} To define the conjugate momenta in the
Hamiltonian formalism, it is necessary to integrate by parts the term
{$g^{-1}\partial^{2}g$} and change it to the form $\partial g\partial g.$
These surface terms, which turned out to be very important, are canceled by
modifying the Euclidean action to {%
\[
I=-\frac{1}{16\pi}%
{\displaystyle\int\limits_{M}}
d^{4}x\sqrt{g}R-\frac{1}{8\pi}%
{\displaystyle\int\limits_{\partial M}}
d^{3}x\sqrt{h}K,
\]
} where {$\partial M$} is the boundary of {$M$, $h_{ab}$ }is the induced
metric on {$\partial M$} and {$K$ }is the trace of the second fundamental form
on $\partial M.$ Notice that there is a relative factor of $2$ between the two
terms, and \ that the boundary term has to be completely fixed. This is a
delicate fine tuning and is not determined by any symmetry, but only by the
consistency requirement. There is no known symmetry that predicts this
combination and it is always added by hand \cite{GH}. In contrast we can
compute the spectral action for manifolds with boundary. The hermiticity of
the Dirac operator {
\[
\left(  \psi\right\vert D\psi)=\left(  D\psi\right\vert \psi),
\]
} is satisfied provided that {$\pi_{-}\psi|_{\partial M}=0$} where {$\pi
_{-}=\frac{1}{2}\left(  1-\chi\right)  $ }is a projection operator on
$\partial M$ with $\chi^{2}=1.$ To compute the spectral action for manifolds
with boundary we have to specify the condition {$\pi_{-}D\psi|_{\partial M}%
=0$.} The result of the computation gives the remarkable result that the
Gibbons-Hawking boundary term is generated without any fine tuning
\cite{boundary}. Adding matter interactions, does not alter the relative sign
and coefficients of these two terms, even when higher orders are included. The
Dirac operator for a product space such as that of the standard model, must
now be taken to be of the form {
\[
D=D_{1}\otimes\gamma_{F}+1\otimes D_{F},
\]
instead of
\[
D=D_{1}\otimes1+\gamma_{5}\otimes D_{F},
\]
} because $\gamma_{5}$ does not anticommute with $D_{1}$ on $\partial M.$

\section{\bigskip Dilaton and the dynamical generation of scale}

Replacing the cutoff scale $\Lambda$ in the spectral action, replacing
{$f(\frac{D^{2}}{\Lambda^{2}})$} by {$f(P)$} where {$P=e^{-\phi}D^{2}e^{-\phi
}$} modifies the spectral action with dilaton dependence to the form
\cite{dilaton} {
\[
\hbox{Tr }F(P)\simeq%
{\displaystyle\sum\limits_{n=0}^{6}}
f_{4-n}%
{\displaystyle\int}
d^{4}x\sqrt{g}e^{\left(  4-n\right)  \phi}a_{n}\left(  x,D^{2}\right)  .
\]
} One can then show that the dilaton dependence almost disappears from the
action if one rescales the fields according to {
\begin{align*}
G_{\mu\nu}  &  =e^{2\phi}g_{\mu\nu}\\
H^{\prime}  &  =e^{-\phi}H\\
\psi^{\prime}  &  =e^{-\frac{3}{2}\phi}\psi.
\end{align*}
} With this rescaling one finds the result that the spectral action {
\[
I\left(  g_{\mu\nu},H,\psi,\phi\right)  =I\left(  G_{\mu\nu},H^{\prime}%
,\psi^{\prime},\phi=0\right)  +\frac{24f_{2}}{\pi^{2}}%
{\displaystyle\int}
d^{4}x\sqrt{G}G^{\mu\nu}\partial_{\mu}\phi\partial_{\nu}\phi
\]
} is scale invariant (independent of the dilaton field) except for the kinetic
energy of the dilaton field $\phi.$ The dilaton field has no potential at the
classical level. It acquires a {Coleman-Weinberg potential} \cite{CW} through
quantum corrections, and thus a vev and a very small mass. \cite{Bubu}. The
Higgs sector in this case becomes identical with the {Randall-Sundrum model
\cite{RS}}. In that model there are two branes in a five dimensional space,
one located at {$x_{5}=0$} representing the invisible sector, and \ another
located at {$x_{5}=\pi r_{c},$} the visible sector. The physical masses are
set by the symmetry breaking scale {$v=v_{0}e^{-kr_{c}\pi}$} so that
{$m=m_{0}e^{-kr_{c}\pi}$.} If the bare symmetry breaking scale is taken at
{$m_{0}\sim10^{19}$} Gev, then by taking {$kr_{c}\pi=10$} one gets the
low-energy mass scale $m\sim10^{2}$ Gev. It is not surprising that the
{Randall-Sundrum} scenario is naturally incorporated in the noncommutative
geometric model \cite{CF}, \cite{Lizzi}, because intuitively one can think of
the discrete space as providing the different right-handed and left-handed
brane sectors.

\section{{{Speculations on the Structure of the Noncommutative Space and
Quantum Gravity}}}

The small deviation from experimental results of the predictions of the
standard model derived from the spectral action can have the following
interpretation. This is an indication that the basic assumption we made about
space-time as a product of a continuous four dimensional manifold times a
discrete space breaks down at energies just below the unification (Planck)
scale. This will lead us to postulate that at Planckian energies, the
structure of space time becomes noncommutative in a nontrivial way, which will
change in an intrinsic way the particle spectrum. On the other hand, the
encouraging results we obtained about the unique prediction of the spectrum of
the standard model, the determination of the gauge group and for particle
representations, can be taken as a guide that the true geometry should
reproduce at lower energies, the product structure we assumed. The starting
point is to look for a noncommutative space whose KO-dimension is ten (mod 8)
and whose metric dimension is dictated by the growth of eigenvalues of the
Dirac operator to be four. \ A good starting point would be to mesh in a
smooth manner the four-dimensional manifold with the discrete space
$M_{2}\left(  \mathbb{H}\right)  \oplus M_{4}\left(  \mathbb{C}\right)  .$ The
appearance of $4\times4$ matrices and their relation to a four-dimensional
space-time may not be accidental. In particular, we can define the
four-dimensional manifold through the following data. The $C^{\ast}$ algebra
is generated by $M_{2}\left(  \mathbb{H}\right)  $ and a projection
$e=e^{2}=e^{\ast}$ such that \cite{Alain}
\begin{align*}
\left\langle e-\frac{1}{2}\right\rangle  &  =0\\
\left\langle \left(  e-\frac{1}{2}\right)  \left[  D,e\right]  ^{2n}%
\right\rangle  &  =\left\{
\begin{array}
[c]{c}%
0,\qquad n=0,1\\
\gamma,\qquad n=2
\end{array}
\right\}  ,\qquad
\end{align*}
where $\gamma$ is the chirality operator satisfying
\[
\gamma^{2}=\gamma,\qquad\gamma=\gamma^{\ast},\qquad\gamma e=e\gamma,\qquad
D\gamma=-\gamma D
\]
The constraint on $e$ forces it to be of the form
\[
e=\left(
\begin{array}
[c]{cccc}%
\frac{1}{2}+t & 0 & \alpha & \beta\\
0 & \frac{1}{2}+t & -\beta^{\ast} & \alpha^{\ast}\\
\alpha^{\ast} & -\beta & \frac{1}{2}-t & 0\\
\beta^{\ast} & \alpha & 0 & \frac{1}{2}-t
\end{array}
\right)
\]
where $t,\alpha,\alpha^{\ast},\beta$ and $\beta^{\ast}$ all commute and
satisfy the relation%
\[
t^{2}+\left\vert \alpha\right\vert ^{2}+\left\vert \beta\right\vert ^{2}%
=\frac{1}{4}.
\]
One can then check that $\mathcal{A}=C\left(  S^{4}\right)  .$ The
differential constraints are then satisfied by any Riemannian structure with a
given volume form on $S^{4}.$ This space can be deformed by considering the
algebra to be generated by $M_{4}\left(  \mathbb{C}\right)  $ and $e$ where
\cite{CL}
\[
e=\left(
\begin{array}
[c]{cc}%
q_{11} & q_{12}\\
q_{21} & q_{22}%
\end{array}
\right)
\]
and each $q$ is a $2\times2$ matrix of the form
\[
q=\left(
\begin{array}
[c]{cc}%
\alpha & \beta\\
-\lambda\beta & \alpha^{\ast}%
\end{array}
\right)
\]
In this case the projection constraints imply
\[
e=\left(
\begin{array}
[c]{cccc}%
\frac{1}{2}+t & 0 & \alpha & \beta\\
0 & \frac{1}{2}+t & -\lambda\beta^{\ast} & \alpha^{\ast}\\
\alpha^{\ast} & -\overline{\lambda}\beta & \frac{1}{2}-t & 0\\
\beta^{\ast} & \alpha & 0 & \frac{1}{2}-t
\end{array}
\right)
\]
satisfying%
\[
\alpha\alpha^{\ast}=\alpha^{\ast}\alpha,\quad\beta\beta^{\ast}=\beta^{\ast
}\beta,\quad\alpha\beta=\lambda\beta\alpha,\quad\alpha^{\ast}\beta
=\overline{\lambda}\beta\alpha
\]
giving rise to deformed $S^{4}.$

The idea now is to define the noncommutative space by marrying the concept of
generating a manifold as instantonic solution of a set of equations, and to
blend these with the finite space. We will report on this in the future.

\section{Conclusions}

We summarize the main assumptions made:

\begin{itemize}
\item Space-time is a product of a continuous four-dimensional manifold times
a finite space.

\item One of the algebras $M_{4}\left(  \mathbb{C}\right)  $ is subject to
symplectic symmetry reducing it to $M_{2}\left(  \mathbb{H}\right)  .$

\item The commutator of the Dirac operator with the center of the algebra is
non trivial $\left[  D,Z\left(  \mathcal{A}\right)  \right]  $ $\neq0.$

\item The unitary algebra $\mathcal{U}\left(  \mathcal{A}\right)  $ is
restricted to $\mathcal{SU}\left(  \mathcal{A}\right)  .$
\end{itemize}

\emph{These give rise to the following results:}

\begin{itemize}
\item The number of fundamental fermions is $16.$

\item The algebra of the finite space is $\mathbb{C}\oplus\mathbb{H}\oplus
M_{3}\left(  \mathbb{C}\right)  .$

\item The correct representations of the fermions with respect to $SU(3)\times
SU(2)\times U(1)$ are derived.

\item The Higgs doublet appears as part of the inner fluctuations of the
metric, and spontaneous symmetry breaking mechanism appears naturally with the
negative mass term without any tuning.

\item Mass of the top quark of around $179$ Gev.

\item See-saw mechanism to give very light left-handed neutrinos.
\end{itemize}

\emph{The following problems are encountered: }

\begin{itemize}
\item The unification of the gauge couplings with each other and with Newton
constant do not meet at one point which is expected to be one order below the
Planck scale.

\item Mass of the Higgs field of around $170$ Gev. This however, depends on
the value of the gauge couplings at the unification scale, which is very uncertain.

\item No new particles besides those of the Standard Model. This will be
\ problemetic if new physics is observed at LHC.

\item No Explanation of the number of generations.

\item No constraints on the values of the Yukawa couplings which are the
non-zero entries in the Dirac operator of the finite space.
\end{itemize}

\emph{From these results we can deduce the following: }

\begin{itemize}
\item It is necessary to include the higher order corrections to the spectral
action using a convergent series for the heat kernel expansion. This step is
now done, and shows clearly that the corrections cannot be ignored if the
spectral function deviates even slightly from the cut-off function. What
remains to be done is to input these corrections into the RG equations and
prove that this mechanism does produce gauge couplings unification, and thus
will enable us to get an accurate prediction for the Higgs mass.

\item To get an insight on the problem of quantum gravity, it is essential to
find the noncommutative space whose limit is the product $M_{4}\times F.$ We
speculated that this could be done by adopting the strategy of generating a
continuous manifold through instantonic solutions of algebraic and
differential constraints. This step has to be elaborated on and we must
construct in detail the structure of such a space, to study its properties at
the Planck scale and to show that the usual space-time can be recovered from
the geometry of a non-trivial noncommutative space.

\item The results obtained so far are very encouraging and we hope to report
on future positive developments.
\end{itemize}

\begin{acknowledgement}
This is supported in part by the Arab Fund for Social and Economic Development
and by NSF grant Phys-0653300.
\end{acknowledgement}

\end{document}